\documentclass[twocolumn,showpacs,preprintnumbers,amsmath,amssymb]{revtex4}
\usepackage{graphicx}
\usepackage{dcolumn}
\usepackage{bm}
\usepackage{times}
\usepackage[colorlinks,citecolor=blue,linkcolor=red]{hyperref}
\usepackage{color}
\usepackage{comment}
\usepackage{mathrsfs}

\begin{document}

	\title{Emergent quantum probability from full quantum dynamics and the role of energy conservation}
	
	\author{Chen Wang}
	%\email{wangchen@zjnu.cn}
	\affiliation{Department of Physics, Zhejiang Normal University, Jinhua, Zhejiang 321004, China}
    \author{Jincheng Lu}
    \affiliation{Jiangsu Key Laboratory of Micro and Nano Heat Fluid Flow Technology and Energy Application, School of Physical Science and Technology,
Suzhou University of Science and Technology, Suzhou, 215009, China}
	\author{Jian-Hua Jiang}
	\email{joejhjiang@hotmail.com}
	\affiliation{Suzhou Institute for Advanced Research, University of Science and Technology of China, Suzhou, 215123, China}

	\date{\today}

	\begin{abstract}
We propose and study a toy model for the quantum measurements that yield the Born's rule of quantum probability. In this model, the electrons interact with local photon modes and the photon modes are dissipatively coupled with local photon reservoirs. We treat the interactions of the electrons and photons with full quantum mechanical description, while the dissipative dynamics of the photon modes are treated via the Lindblad master equation. By assigning double quantum dot setup for the electrons coupling with local photons and photonic reservoirs, we show that the Born's rule of quantum probability can emerge directly from microscopic quantum dynamics. We further discuss how the microscopic quantities such as the electron-photon couplings, detuning, and photon dissipation rate determine the quantum dynamics. Surprisingly, in the infinite long time measurement limit, the energy conservation already dictates the emergence of the Born's rule of quantum probability. For finite-time measurement, the local photon dissipation rate determines the characteristic time-scale for the completion of the measurement, while other microscopic quantities affect the measurement dynamics. Therefore, in genuine measurements, the measured probability is determined by both the local devices and the quantum mechanical wavefunction.	\end{abstract}

%\pacs{03.65.Ta,42.50.Ct,03.65.Yz,74.25.Fy}
%03.65.Ta Foundations of quantum mechanics; measurement theory
%03.65.Yz Decoherence; open systems; quantum statistical methods
%42.50.Ct Quantum description of interaction of light and matter; related experiments
%42.50.-p Quantum optics
%74.25.Fy Transport properties (electric and thermal conductivity, thermoelectric effects, etc.)
	\maketitle

	\section{Introduction}
The emergence of probability is one of the key features in quantum mechanics that has been puzzling since the birth of quantum mechanics~\cite{sw1989ap,sw1989prl}. In 1926 Max Born proposed his theory where quantum states are interpreted as probability waves~\cite{mborn1926zp,mborn1954nl}. This probability wave theory connects the probability in quantum measurements with the well-known quantum probability, i.e., the square of the absolute value of the wavefunction. This interpretation, though gives ease in using quantum theory to compute experimental observables, it never fully satisfied the questioning minds in the past century~\cite{usinha2010science,abassi2013rmp,pshadbolt2014np,mop2021prl}. In particular, how could a wave be probabilistic when the wave equation (i.e., the Schr\"odinger equation) is deterministic? What if the connection between the quantum wavefunction and the measured probability is an emergent phenomenon, i.e., it emerges from the quantum dynamics of the measurement process instead of a direct connection as interpreted by Born?

In the past decades, it has been accepted that probability emerges in quantum mechanics due to the interaction between the quantum system and the measurement instrument~\cite{jaw1984book,hmwiseman2011book}. The instrument generally contains very large degrees of freedom and serves as an environment interacting with the quantum system. Such interaction causes quantum decoherence that leads to the loss of certainty and the emergence of quantum probability~\cite{whz2003prl,whz2005pra,ms2005fop,um2004ijqi}. Several models have been proposed to explore the underlying physics of quantum measurement~\cite{aea2013rmp}, e.g., the Coleman-Hepp model~\cite{bs2001book}, the Gaveau-Schulman model~\cite{gb1990jsp}, and the Curie-Weiss model~\cite{aea2003epl}. In particular, the Curie-Weiss model consisting of a spin 1/2 measured by coupled-spins interacting with one phonon reservoir, was used to elucidate the practical measurement in spin systems~\cite{aea2013rmp}. However, a simple microscopic model of quantum measurements which can demonstrate the emergence of Born's quantum probability from a full quantum mechanical description is still helpful in understanding the underlying physics of quantum measurements.

Here, we propose such a model by considering the detection of electron's probability in a double quantum dot setup via the local electron-photon interactions~\cite{mrd2011prl,yyliu2014prl,xmi2017science,trh2018prl}. In this model, electrons interacts with local photon modes and the photon modes dissipate their energy into local photonic reservoirs. We treat the electron-photon interaction in a fully quantum mechanical framework and describe the photon dissipation via the Lindblad master equation~\cite{uweiss2012book,hpb2002book}. We propose to measure the quantum system by monitoring the photonic energy dissipation. By solving the quantum dynamic equations, we find that in the infinite long time limit, the photonic energy dissipation into the reservoir is equal to the input electron energy. Since the latter is proportional to the square of the absolute value of the wavefunction, the energy conservation already dictates that the measured quantum probability is equal to the Born's rule of quantum probability. For finite time measurements, the local photon dissipation rate determines the characteristic timescale for the completion of the measurement, while other microscopic quantities affect the measurement dynamics. We further discuss how electron-photon detuning, strong electron-photon interaction, and finite reservoir temperature affect the measurement dynamics. Our work reveals the underlying microscopic quantum physics of quantum measurements and can serve as a prototype model to study the origin of the Born's rule on the quantum probability.
Throughout this work, we set Planck constant $\hbar\equiv1$.

	%==========================================
	\begin{figure}[tbp]
		\begin{center}
			%\vspace{-1.0cm}
			\includegraphics[scale=0.6]{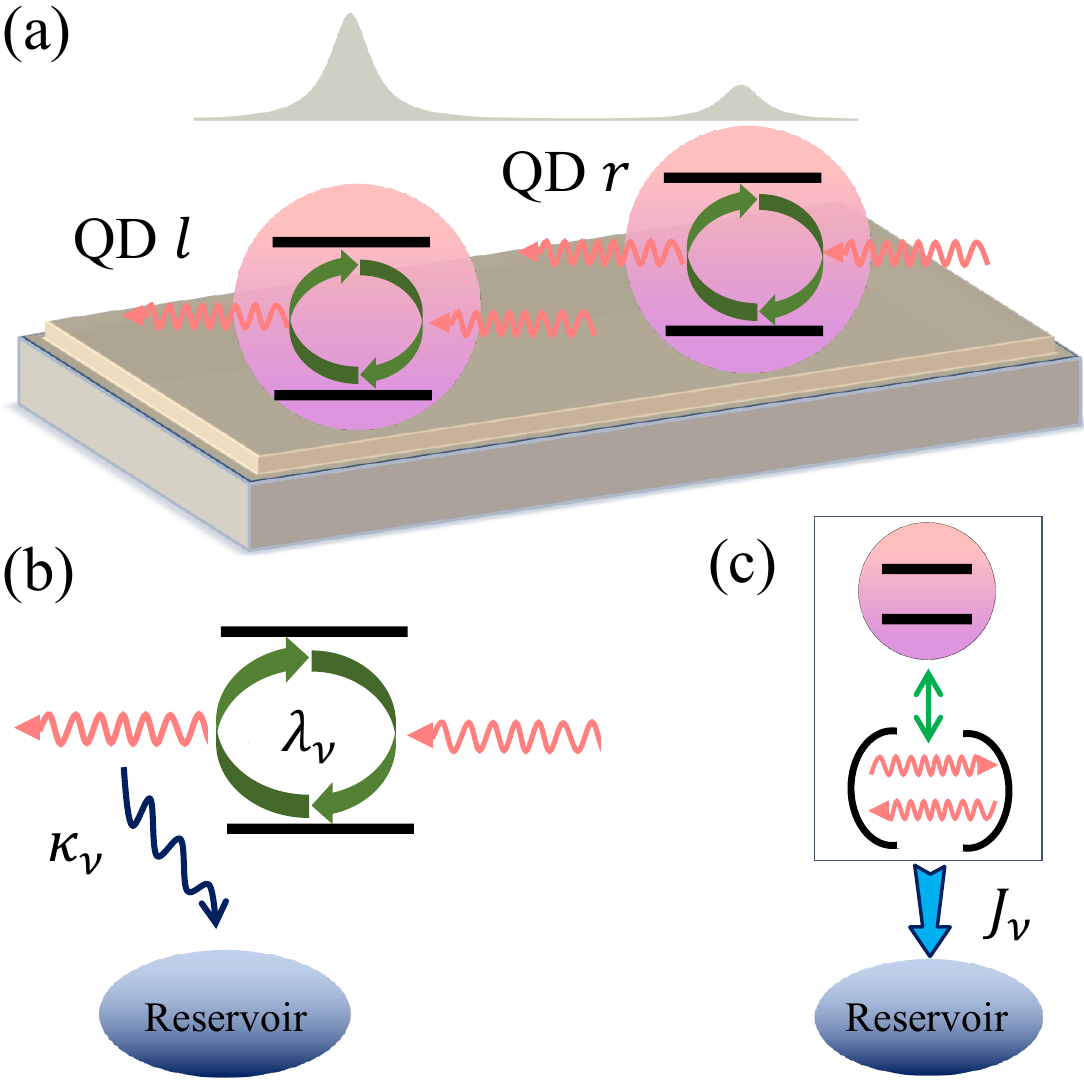}
			%\vspace{-1.0cm}
			\caption{(Color online)
				(a) The schematic illustration of the electron-photon model at Eq.~(\ref{hs1}).
				It is composed of two individual quantum dots, each owning one electron and interacting with single-mode photonic resonator.
				(b) The description of quantum probability readout from the quantum dots at Eq.~(\ref{lindbladv}) via the electron-photon interaction $\lambda_v$ and the photon damping process $\kappa_v$.
				(c) The microscopic energy conservation process during the measurement process at Eq.~(\ref{je1}), i.e., the energy loss from the electron-photon hybrid system is identical with the gain into the reservoir, which is characterized as the energy current $J_v$.
			}~\label{fig0}
		\end{center}
	\end{figure}
	%%==========================================

	\section{Measurement setup}
	We begin by introducing a generic quantum system that consists of two individual quantum dots interacting with one photonic resonator~\cite{syzhu1988pla,vva2012prl,pstegman2022acs}, to describe the probability measurement of the electron wavefunction through the electron-photon interactions, where the photonic resonators may be treated as the  probability detectors~\cite{rl2020prx,ea2021nature}. Then, we analyze the dissipative dynamics of the electron-photon interacting model to practically read out information on the quantum probability of the electron wavefunction in each quantum dot. We note that the measurement description of the electron-photon interaction and photon dissipation is fully quantum.
	
	\subsection{Electron-photon  Model}
	The electron-photon system is modeled as
	two  quantum dots each owning single electron, separately coupled to two photonic  resonators, shown in Fig.~\ref{fig0}(a).
	The Hamiltonian is described as $H_s=\sum_{v={l},{r}}H_{s,v}$.
	The $v$-th system component is specified as
	\begin{eqnarray}~\label{hs1}
		H_{s,v}&=&\sum_{j=e,g}E_{j,v}|j_{v}{\rangle}{\langle}j_v|+\omega_{0,v} a^\dagger_v a_v\\
		&&+\lambda_v(a^\dagger_v|g_{v}{\rangle}{\langle}e_v| +a_v |e_{v}{\rangle}{\langle}g_v|),\nonumber
	\end{eqnarray}
	where $|e_{v}{\rangle}$ and $|g_{v}{\rangle}$ are the excited and ground states
	with the corresponding occupation energies $E_{v,e}$ and $E_{v,g}$ in the $v$-th quantum dot,
	$a^\dag_v~(a_v)$ is the creating (annihilating) operator of the  photonic resonator with the photon energy $\omega_{0,v}$,
	$\lambda_v$ denotes the electron-photon coupling strength.
	In the following, we set $E_{e,{l}}=E_{e,{r}}=0$ without losing any generality.
	
	Considering $H_{s,v}$
	in the subspace $\{|n_v-1,e_{v}{\rangle},|n_v,g_{v}{\rangle}\}$
	with $n_v$ being the photon excitation number,
	it is straightforward to find that the $n$-th eigenvalues of the electron-photon model are given by
	$E_{n_v,\pm}=\omega_{0,v}n_v+\frac{E_{g,v}-\omega_{0,v}}{2}\pm\sqrt{(\frac{\omega_{0,v}+E_{g,v}}{2})^2+\lambda^2_vn_v}$,
	and the eigenvectors are given by
	\begin{subequations}
		\begin{align}
			&	\vert \psi_{n_v,+}\rangle =\cos{\frac{\theta_{n_v}}{2}}\vert n_v,g_{v}\rangle +\sin{\frac{\theta_{n_v}}{2}}\vert n_v-1,e_{v}\rangle, \\
			&  \vert \psi_{n_v,-}\rangle =-\sin{\frac{\theta_{n_v}}{2}}\vert n_v,g_{v}\rangle +\cos{\frac{\theta_{n_v}}{2}}\vert n_v-1,e_{v}\rangle ,
		\end{align}
	\end{subequations}
	with $\tan\theta_{n_v}=2\lambda_v\sqrt{n_v}/(\omega_{0,v}+E_{g,v})~(n_v{\ge}1)$.
	Moreover, we define the effective ground state
	$|\psi_{0_v}{\rangle}=|0,g_{v}{\rangle}$
	and the corresponding energy $E_{0,v}=E_{g,v}$.
	It need note that the real bare ground state actually is $|0, g_v{\rangle}$. However, this state will never take participate
	into the dissipative dynamics of the dot-photon system.
	
	\subsection{Dynamics}
	We consider master dynamics of the electron-photon  model with
	weak photon dissipation~\cite{uweiss2012book,hpb2002book,pstegman2022acs}
	\begin{eqnarray}~\label{lindblad}
		\frac{d \rho_s}{dt}&=&-i[H_s,\rho_s]\\
		&& -\sum_{v={l},{r}}\frac{\kappa_v}{2}[a^\dagger_v a_v \rho_s +\rho_s a^\dagger_v a_v -2a_v\rho_s a^\dagger_v],\nonumber
	\end{eqnarray}
	where $\kappa_v$ means the dissipation strength of  photons from the $v$-th resonator into the corresponding zero-temperature environment (e.g., the $v$-th photon counter).
	   Here, the dissipation of quantum dots is assumed negligible compared to the photon damping strength.
	Then, we select the initial state of the electron-photon model.
	There exists one electron in each quantum dot,
	i.e.,
	$|\phi_{e}(0)=C_{l}|e_{{l}},g_{{r}}{\rangle}
	+C_{r}|g_{{l}},e_{{r}}{\rangle}$,
	and resonators are initially both in vacuum before the procedures of the probability readout.
	Hence, the initial wavefunction is generally described as
	\begin{eqnarray}~\label{is1}
		|\phi(0){\rangle}&=&C_{l}|0_{l},e_{{l}}{\rangle}{\otimes}
		|0_{r},g_{{r}}{\rangle}
		+C_{r}|0_{l},g_{{l}}{\rangle}{\otimes}|0_{r},e_{{r}}{\rangle},
	\end{eqnarray}
	where $C_v$ is the wavefunction coefficient.

We focus on the individually measuring the quantum probability in the local dots~\cite{man2002book,jcorler2020prl}.
	Consequently, the initial reduced density matrix of the $v$-th system component is obtained as
	\begin{eqnarray}~\label{is2}
		\rho_{v}(0)=\mathbf{P}_v|0_v,e_{v}{\rangle}{\langle}0_v,e_{v}|
		+\mathbf{P}_{\overline{v}}|0_{v},g_{v}{\rangle}{\langle}0_{v},g_{v}|,
	\end{eqnarray}
where the Born's quantum probabilities are expressed as
\begin{eqnarray}~\label{bqp1}
\mathbf{P}_v=|C_v|^2,
\end{eqnarray}
with ${\overline{v}}={r}$ for $v={l}$ and  ${\overline{v}}={l}$ for $v={r}$.
	Hence, the dissipative dynamics of the $v$-th component is expressed as
	\begin{eqnarray}~\label{lindbladv}
		\frac{d \rho_v}{dt}=-i[H_{s,v},\rho_v]-\frac{\kappa_v}{2}[a^\dagger_v a_v \rho_v +\rho_v a^\dagger_v a_v -2a_v\rho_v a^\dagger_v],
	\end{eqnarray}
	with $\rho_{{l}({r})}
	=\textrm{Tr}_{{r}({l})}\{\rho_s\}$, which is also exhibited in Fig.~\ref{fig0}(b).
 	Moreover, the dynamics is constraint to the lowest three eigenstates,
	i.e.,
	$\vert \psi_{1_v,+}\rangle$, $\vert \psi_{1_v,-}\rangle$,
	and $|\psi_{0_v}{\rangle}$.
	And it is straightforward to show the inverse relations
	$|0_v,e_{v}{\rangle}=(\sin{\frac{\theta_v}{2}}\vert \psi_{1_v,+}\rangle +\cos{\frac{\theta_v}{2}}\vert \psi_{1_v,-} \rangle)$
	and
	$|1_v,g_{v}{\rangle}=(\cos{\frac{\theta_v}{2}}\vert \psi_{1_v,+}\rangle -\sin{\frac{\theta_v}{2}}\vert \psi_{1_v,-} \rangle)$,
	with $\tan{\theta_v}={2\lambda_v}/{(\omega_{0,v}+E_{g,v})}$.

	\section{Quantum probability Measurements}
	In this section, we investigate the probability measurement at weak and strong electron-photon couplings and finite temperature reservoir, respectively.
	
	\subsection{Measurement with weak electron-photon coupling}
	We first study the  probability measurement based on the quantum dots and  photonic resonators in the long time evolution limit,
	i.e., complete measurement.
	Then, we further analyze the finite time behavior of the emitted photons, i.e., finite time measurement.
	Finally, we briefly discuss the influence of finite temperature reservoir on measured probability.
	
	\subsubsection{Complete measurement}
	We investigate the probability measurement from the energy perspective in the double quantum dot setup via the interaction with local photon resonators,
	with the initial state at Eq.~(\ref{is1}).
	Specifically, the energy flow from the $v$-th system component
	into the dissipative reservoir (e.g., the $v$-th photon counter) can be defined as
	\begin{eqnarray}~\label{je1}
		J_{v}=-d{\langle}H_{s,v}{\rangle}(t)/dt,
	\end{eqnarray}
	which quantifies the  energy loss of the $v$-th component, shown in Fig.~\ref{fig0}(c).
	Meanwhile, in the eigenspace of $H_{s,v}$,
	the expression of the energy flow can be reexpressed as
	$J_v=-[E_{+,v}(dP_{v,+}(t)/dt)+E_{-,v}(dP_{v,-}(t)/dt)]$,
	with  the density matrix elements denoting
	$P_{v,\pm}(t)=\langle \psi_{1_v,\pm}\vert \rho_v(t) \vert \psi_{1_v,\pm}\rangle$
	and $P_{v,+-}(t)=\langle \psi_{1_v,+}\vert \rho_v(t) \vert \psi_{1_v,-}\rangle$,
	and energies given by
	\begin{eqnarray}~\label{Epm}
		E_{+(-),v}=\Big[(\omega_{0,v}-E_{g,v}){\pm}
		\sqrt{(\omega_{0,v}+E_{g,v})^2+4\lambda^2_v}\Big]/2,
	\end{eqnarray}
	which characterizes the microscopic energy transport processes, e.g.,
	$E_{+,v}[-dP_{v,+}(t)/dt]$ characterizing the relaxation of the system from
	the channel $|\psi_{1_v,+}{\rangle}$
	to $|\psi_{0_v}{\rangle}$ by releasing the energy $E_{+,v}$ and one excitation particle.
	After sufficient long time evolution,
	it is straightforward to find that the total energy loss of the quantum system is obtained as
	$\int^\infty_0dt[-d{\langle}H_v{\rangle}(t)/dt]
	=-E_{g,v}|C_v|^2$.
	Simultaneously, the microscopic transferred energy to the reservoir
	is given by integrating the current
\begin{eqnarray}	
-\int^\infty_0dt\Big[E_{+,v}\frac{dP_{v,+}(t)}{dt}+E_{-,v}\frac{dP_{v,-}(t)}{dt}\Big]
	=-E_{g,v}|C_v|^2.
\end{eqnarray}
	This clearly demonstrates the microscopic energy conservation, i.e., the energy of the electron
	in the $v$-th dot is fully dissipated into the corresponding reservoir via two eigen-channels.

	Furthermore, it is interesting to find that
	Based on the master equation, the energy flow
	at Eq.~(\ref{je1}) can be reexpressed as
	\begin{eqnarray}~\label{je2}
		J_v&=&E_{+,v}\kappa_v[\cos^2\frac{\theta_v}{2}P_{v,+}(t)
		-\cos\frac{\theta_v}{2}\sin\frac{\theta_v}{2}\textrm{Re}\{P_{v,+-}\}]\nonumber\\
		&&+E_{-,v}\kappa_v[\sin^2\frac{\theta_v}{2}P_{v,-}(t)
		-\cos\frac{\theta_v}{2}\sin\frac{\theta_v}{2}\textrm{Re}\{P_{v,+-}\}].\nonumber\\
	\end{eqnarray}
	Both the populations
	and coherence terms contribute to the energy flow.
	Moreover, the dynamics of the photon number in the $v$-th resonator is described as
	\begin{eqnarray}~\label{nt1}
		\langle a^\dagger_v a_v\rangle(t) &=&\cos^2{\frac{\theta_v}{2}} P_{v,+}(t) + \sin^2{\frac{\theta_v}{2}}P_{v,-}(t)\\
		&&-2\cos{\frac{\theta_v}{2}}\sin{\frac{\theta_v}{2}}\textbf{Re}[P_{v,+-}(t)].\nonumber
	\end{eqnarray}
	Then, we consider the dynamical equation at Eq.~(\ref{lindbladv})
	and integrate the photonic energy flow $\int^\infty_0J_vdt$ at Eq.~(\ref{je2}), which is connected with the countered photon numbers (see Appendix for the details).
By defining the measured quantum probability
\begin{eqnarray}~\label{mqp1}
\mathscr{P}_v(t) {\equiv} \kappa_v\int ^{\infty}_{0}dt\langle a^\dagger_v a_v\rangle(t),
\end{eqnarray}
 the complete quantum probability measurement inherited from microscopic energy conservation is established as
	\begin{eqnarray}~\label{cm1}
		\mathscr{P}_v(\infty)=\mathbf{P}_v,
	\end{eqnarray}
with $\mathbf{P}_v=|C_v|^2$.
	This relation denotes that by practically counting the emitted photon number from the individual resonator to the external sink (i.e., the measured probability), we are able to correctly detect the quantum probability of the electron wavefunction in each dot,
	once the dissipation rate $\kappa_v$ is available.
	And the measured probability is irrelevant with the electron-photon system parameters, e.g., $\omega_{0,v}$, $E_{g,v}$,
	and $\lambda_v$.
	Therefore, we conclude that the microscopic energy conservation is able to yield the complete probability measurement.

	Alternatively, from the particle conservation perspective,
i.e., from the excitation particles in the hybrid system to detected photon particles via the photon dissipation processes, we find the dynamics of the ground state population
	($P_{v,0}(t)={\langle}\psi_{0_v}|\rho_v(t)|\psi_{0_v}{\rangle}$) is tightly related with the photon flow, i.e.,
	${dP_{v,0}(t)}/{dt}=\kappa_v\langle a^\dagger_v a_v\rangle(t)$.
	This demonstrates that the probability of electrons transport into the ground state  can be read out via  the emitted photon flows
	$\kappa_v\langle a^\dagger_v a_v\rangle(t)$.
	After the long time evolution the complete probability measurement at Eq.~(\ref{cm1}) can be fully
	recovered,
	considering $P_{v,0}(0)=|C_{\overline{v}}|^2$ and $P_{v,0}(\infty)=1$.

	\subsubsection{Finite time measurement}
	Here, we study the probability measurement at finite time dynamics.
	Since the complete probability measurement holds individually for each electron-photon interacting component, we consider the initial state  at Eq.~(\ref{is1}).
	%$|\phi(0)=|0_L,e_{L,1}\rangle{\otimes}|0_R,e_{R,2}\rangle$.
	It should note that even for such a concise reduced density matrix Eq.~(\ref{is2}),
	the general solution of dissipative dynamics is hard to obtain, due the coupling between diagonal and off-diagonal elements of the electron-photon system.
	While if we include the secular approximation,
	the equations of motion of system density matrix are reduced to
	${dP_{v,+}}/{dt}{\approx}{-\kappa_v}\cos^2 \frac{\theta_v}{2}P_{v,+}$,
	${dP_{v,-}}/{dt}{\approx}{-\kappa_v}\sin^2 \frac{\theta_v}{2}P_{v,-}$,
	and
	${dP_{v,+-}}/{dt}\approx-(i \Lambda_v P_{v,+-}+\frac{\kappa_v}{2}P_{v,+-})$,
	with the energy gap
	$\Lambda_v=\sqrt{({\omega_{0,v}+E_{g,v}})^2+4\lambda^2_v}$
	between $|\psi_{1_v,+}{\rangle}$ and  $|\psi_{1_v,-}{\rangle}$.
	The dynamics of density matrix elements are given by
	$P_{v,+}(t)\approx|C_v|^2\exp(-\kappa_v\cos^2\frac{\theta_v}{2}t)\sin^2\frac{\theta_v}{2}$,
	$P_{v,-}(t)\approx|C_v|^2\exp(-\kappa_v\sin^2\frac{\theta_v}{2}t)\cos^2\frac{\theta_v}{2}$,
	$P_{v,+-}(t)\approx|C_v|^2\exp[-(iE_{v,+-}+\kappa_v/2)t]\cos\frac{\theta_v}{2}\sin\frac{\theta_v}{2}$.
	Then, the accumulated photonic energy base on Eq.~(\ref{je2}) into the dissipative reservoir is approximated as
	
	\begin{eqnarray}
		\int^t_0J_v(s)ds&\approx&|C_v|^2\Big[
		E_{+,v}\sin^2\frac{\theta_v}{2}(1-e^{-(\cos^2\frac{\theta_v}{2})\kappa_vt})\nonumber\\
		&&+E_{-,v}\cos^2\frac{\theta_v}{2}(1-e^{-(\sin^2\frac{\theta_v}{2})\kappa_vt})\nonumber\\
		&&-\kappa_v\frac{\sin^2\theta_v}{2}(E_{+,v}+E_{-,v})\nonumber\\
		&&{\times}
		\textrm{Re}\{\frac{1-e^{-(i\Lambda_v+\frac{\kappa_v}{2})t}}{i\Lambda_v+\frac{\kappa_v}{2}}\}
		\Big].
	\end{eqnarray}	
After long time evolution $t{\rightarrow}\infty$,
	the accumulated photonic energy is given by
	$\int^\infty_0J_v(s)ds{\approx}
	|C_v|^2(E_{+,v}\sin^2\frac{\theta_v}{2}+E_{-,v}\cos^2\frac{\theta_v}{2}{\approx}-E_{g,v}|C_v|^2$
	under the condition $\lambda_v{\gg}\kappa_v$,
	which approaches the total energy dissipated to the reservoir.
	
	Similarly, the measured quantum probability, i.e., the detected photon numbers at Eq.~(\ref{mqp1}), is generally given by
	%\begin{widetext}
		\begin{eqnarray}
			\mathscr{P}_v(t) & \approx&
			|C_v|^2\frac{\sin^2\theta_v}{4}
			\Big[\frac{1-e^{-(\cos^2\frac{\theta_v}{2})\kappa_vt}}{\cos^2\frac{\theta_v}{2}}\nonumber\\
			&&+\frac{1-e^{-(\sin^2\frac{\theta_v}{2})\kappa_vt}}{\sin^2\frac{\theta_v}{2}} - 2\kappa_v\textrm{Re}\Big\{\frac{1- e^{-(i\Lambda_v+\frac{\kappa}{2})t}}{\frac{\kappa}{2}+i\Lambda_v}\Big\}\Big],\nonumber
		\end{eqnarray}
		%\end{widetext}
	In the limit $t{\rightarrow}\infty$, we find that
	$\kappa_v\int^\infty_0\langle a^\dagger_v a_v\rangle(s)ds \approx
	\Big[1
	-\frac{\cos^2\frac{\theta_v}{2}\sin^2\frac{\theta_v}{2}\kappa^2_v}{(\kappa_v/2)^2+E^2_{v,+-}}\Big]$.
	In the coupling regime comparatively stronger than the dissipation strength $\lambda_v{\gg}\kappa_v$,
	we obtain $\kappa_v\int^\infty_0\langle a^\dagger_v a_v\rangle(s)ds \approx |C_v|^2$,
	which consistently recovers the complete probability measurement.
	For convenience, we select system parameters as
	$\omega_{0,v}=\omega_0$,
	$E_{g,v}=E_g$,
	$\lambda_v=\lambda$,
	and the dissipative strength $\kappa_v=\kappa$.

	%==========================================
	\begin{figure}[tbp]
		\begin{center}
			%\vspace{-1.0cm}
			\includegraphics[scale=0.45]{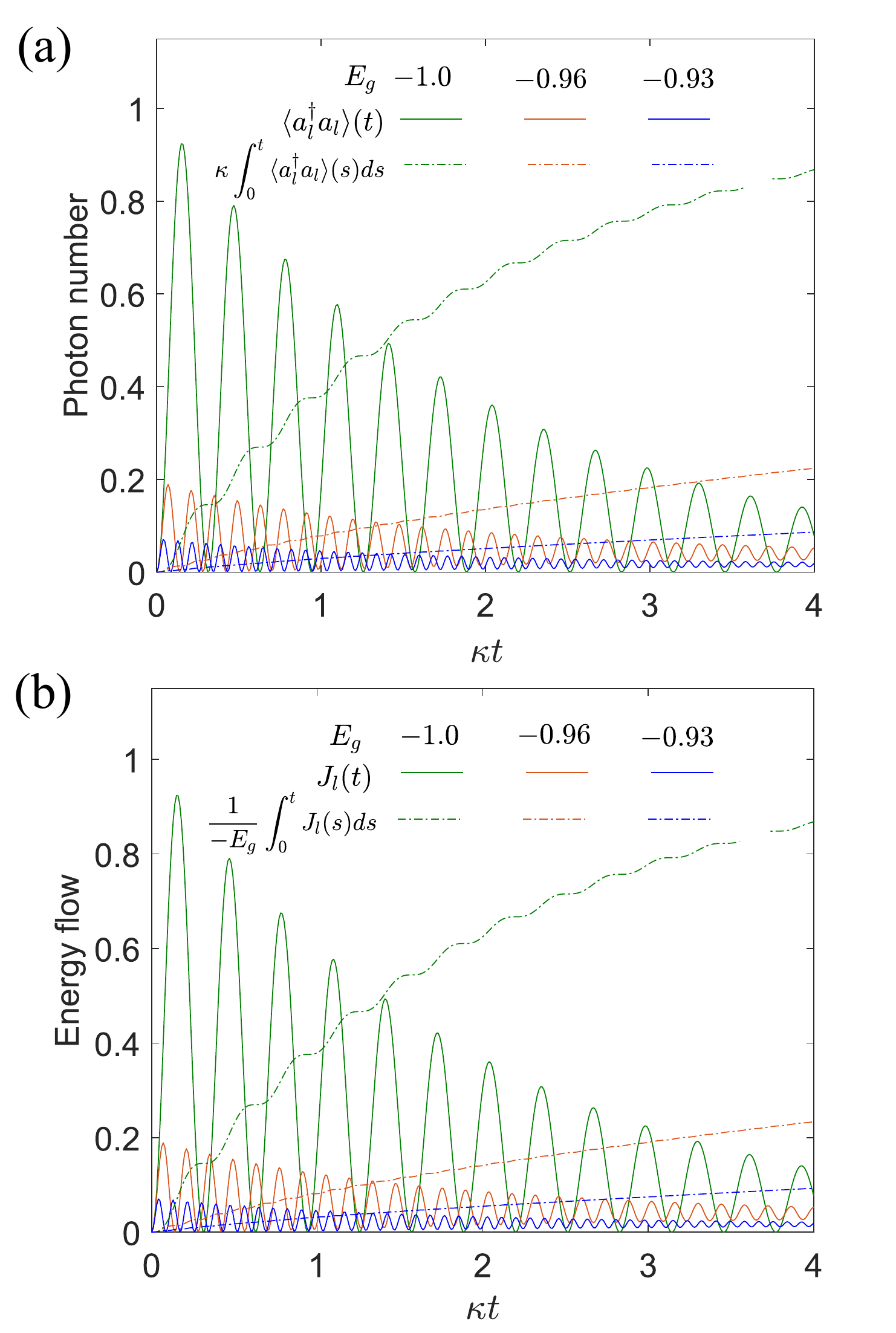}
			%\vspace{-1.0cm}
			\caption{(Color online) The finite time dynamics of (a) photon numbers and (b) energy flow with various quantum dot energies. The initial state is selected as $|\phi(0){\rangle}=|0_l,e_{l}{\rangle}{\otimes}|0_r,g_{r}{\rangle}$. The system parameters are given by $\omega_0=1$,
				$E_{e}=0$,
				$\lambda=0.01$,
				and $\kappa=0.001$.
			}~\label{fig1}
		\end{center}
	\end{figure}
	%%==========================================

	Then, we plot Fig.~\ref{fig1} to study the  finite time dynamics of
	the dissipated energy and
	the photon numbers with $C_l=1$ and $C_r=0$.
	Thus, only the $l$-th part component contributes to the dissipative dynamics.
	In particular at resonance ($E_g=-\omega_0$), it is found that
	the average photon number and
	the photonic energy flow
	are reduced to
	\begin{subequations}
		\begin{align}
			{\langle}a^\dag_{l}{a}_{l}{\rangle}(t) \, \approx& \,  \exp(-{\kappa}t/2)(1-\cos2\lambda{t})/2,~\label{nt}\\
			J_{l}(t) \,  {\approx}\,  &{\kappa}(-E_g)
			\exp(-\kappa{t}/2)(1-\cos2\lambda{t})/2,~\label{je3}
		\end{align}
	\end{subequations}
	which both exhibits periodic damping oscillation behavior with the period $T=\pi/\lambda$.
	Accordingly, the measured quantum probability and dissipated energy  show intermittent increase.
	In particular, the $n$-th ``quasi-step"~\cite{note1} with the amplitude
	\begin{subequations}
		\begin{align}
			\mathscr{P}(nT)
			\,{\approx}\, & [1-e^{-n\pi\kappa/2\lambda}]~\label{icm1},\\
			\frac{1}{-E_{g}}\int^{nT}_0J_{l}(s)ds\, {\approx}\, &
			[1-e^{-n\pi\kappa/2\lambda}],~\label{je4}
		\end{align}
	\end{subequations}
	are shown around the time ${t}=nT$,
	where the photon counter detects the negligible photon flow.
	{From Eqs.~(\ref{icm1}) and (\ref{je4}) it is known that the characteristic measurement number of times can be quantified as $n_{c}{\approx}2\lambda/\kappa$ and the corresponding time is given by
$t_{c}=n_cT{\approx}2\pi/\kappa$. Then the measurement outcome of emitted photons number and energy gradually becomes complete, which approaches the complete measurement results.
While at a given finite measurement time, the effectiveness and systematic error of quantum probability measurement generally rely on the specific properties of the measurement devices, which are typically characterized as $\lambda$ and $\kappa$.
}

	Moreover, we analyze the influence of the bias condition ($E_{g}{\neq}-\omega_{0}$)
	on the dissipative dynamics of photons.
	Both the photon numbers and energy flow
	are monotonically suppressed with the increase of energy bias $|\omega_0+E_g|$
	(though not shown here, such monotonic suppression  also persists in the regime ($-E_g{>}\omega_0$).
	Hence, the fast detection of the complete quantum probability and the underlying conservation  favors the resonant condition.
	
		%==========================================
	\begin{figure}[tbp]
		\begin{center}
			%	\vspace{-3.0cm}
			\includegraphics[scale=0.4]{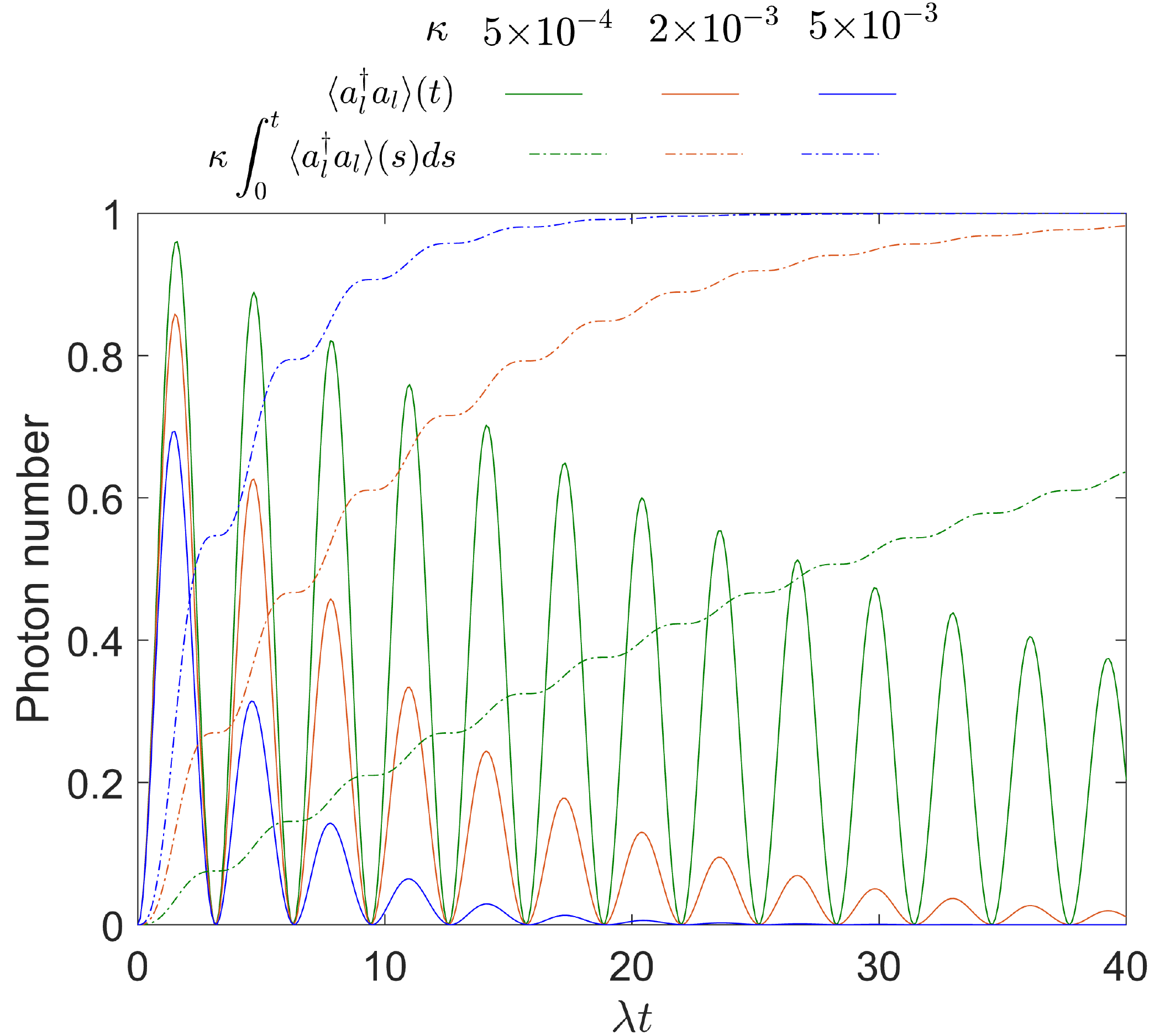}
			%	\vspace{-3.0cm}
			\caption{(Color online) The finite time dynamics of photon number and accumulation
				with various photon dissipation strengths.
				The initial state is the same as Fig.~\ref{fig1}.
				The system parameters are given by $\omega_0=1$,
				$E_e=0$,
				$E_g=-1$,
				and $\lambda=0.01$.
			}~\label{fig2}
		\end{center}
	\end{figure}
	%%==========================================

	We also study the effect of the photon dissipation on the dynamics of photons numbers at resonance in Fig.~\ref{fig2},
	considering the analogous behaviors of energy flow as shown in Eqs.~(\ref{je3})
	and (\ref{je4}).
	For the average photon number, it is shown that the profile is dramatically suppressed with the increase of the dissipation strength $\kappa$,
	whereas the oscillating period is robust against the dissipation process.
	This is consistent with Eq.~(\ref{nt}) and quantified by the characteristic time $2\pi/\kappa$.
	Accordingly, the emitted photon numbers exhibit fast accumulation in the photon counter.
	Therefore, we conclude that enhancing the photon dissipation is helpful for the efficient measurement of the emitted photons and energy.

	%==========================================
	\begin{figure}[tbp]
		\begin{center}
			%\vspace{-3.0cm}
			\includegraphics[scale=0.5]{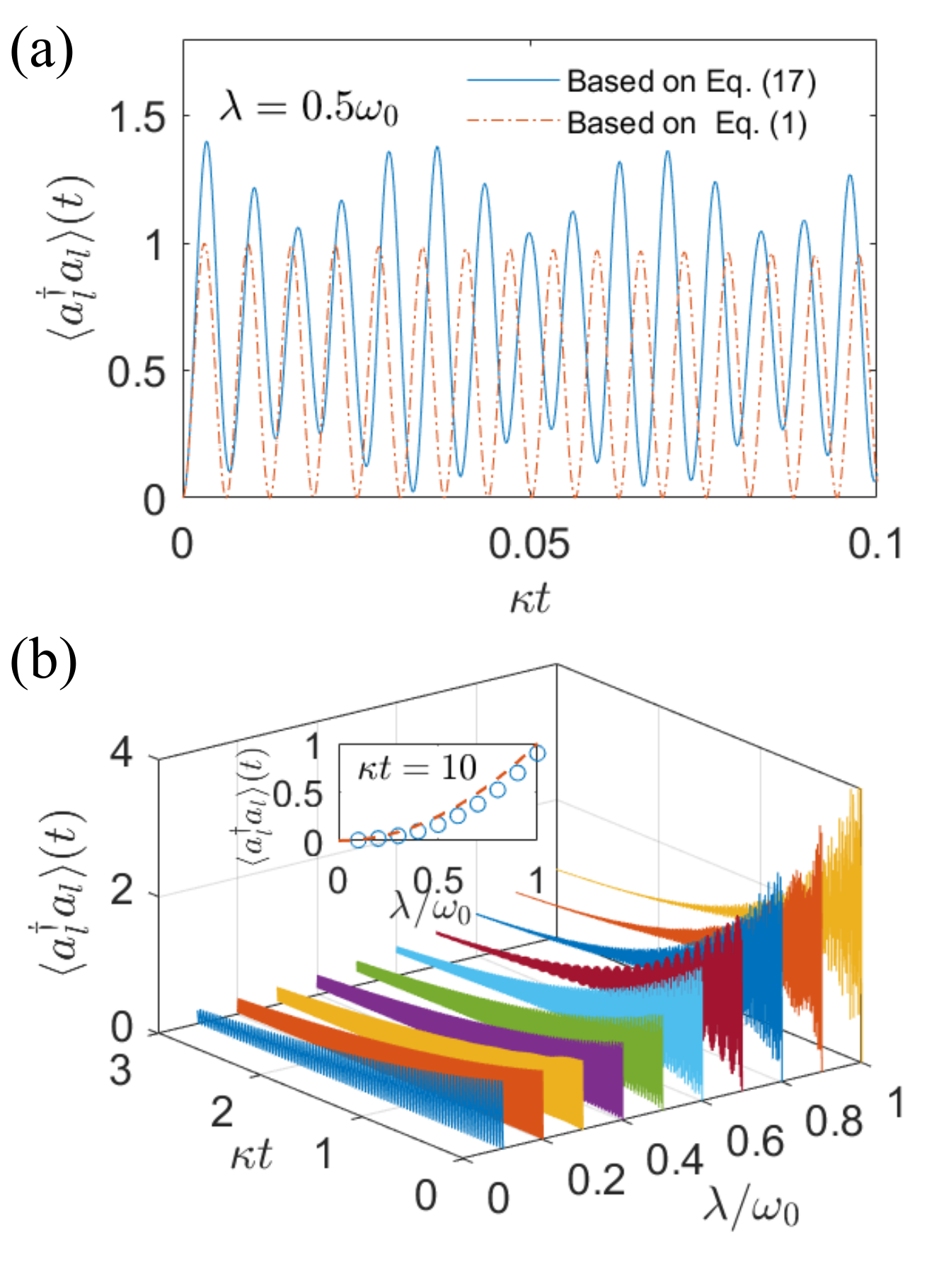}
			%\vspace{-3.0cm}
			\caption{(Color online) The finite time dynamics of photon number with strong electron-photon interaction
				(a) at $\lambda=0.5\omega_0$
				and
				(b) by tuning coupling strength $\lambda$.
				The inset in (b) denotes the photon number as a function of electron-photon coupling strength,
				where the circles and dashed lines are calculated from full numerical calculation at the time ${\kappa}t=10$ and $(\lambda/\omega_0)^2$, respectively.
				The initial state is the same as Fig.~\ref{fig1}.
				The system parameters are given by $\omega_0=1$,
				$E_e=0$,
				and
				$E_g=-1$.
			}~\label{fig3}
		\end{center}
	\end{figure}
	%%==========================================

	\subsection{Measurement with strong electron-photon coupling}
	When the electron-photon coupling becomes strong
	and even comparable with the bare frequency of the optical resonator, the counter-rotating terms of the electron-photon interaction should be considered~\cite{mk2014prb,bka2016prb,jclu2019prb}.
	The system then becomes analogous with the seminal quantum Rabi model~\cite{dbraak2011prl,afk2019nrp,pfd2019rmp},
	is modified to
	\begin{eqnarray}~\label{hs2}
		H_{s,v}&=&\sum_{j=e,g}E_{j,v}|j_{v}{\rangle}{\langle}j_v|+\omega_{0,v} a^\dagger_v a_v\\
		&&+\lambda_v(a^\dagger_v+a_v)(|g_{v}{\rangle}{\langle}e_v|+ |e_{v}{\rangle}{\langle}g_v|),\nonumber
	\end{eqnarray}
	Due to the inclusion of the counter-rotating terms
	$(a^{\dagger}_v|e_{v}{\rangle}{\langle}g_v|+a_v|g_{v}{\rangle}{\langle}e_v|)$, the particle conservation breaks  down,
	whereas the energy conservation is still valid.
	Accordingly, we find that the analytical eigensolution of $H_{s,v}$ is hard to obtain until now, except the solution mapped to the special function via Bargmann algebra~\cite{dbraak2011prl} and
	photonic coherent states~\cite{qhchen2012pra}.
	Here, we include the master equation Eq.~(\ref{lindblad})
	to study the influence of strong electron-photon interaction on probability measurement.

	With the initial state $|\phi(0)=|0_{l},e_{{l}}\rangle{\otimes}|0_{r},g_{{r}}\rangle$,
we first plot  Fig.~\ref{fig3}(a) to study the short-time dissipative dynamics of average photon number
	${\langle}a^\dag_la_l{\rangle}(t)$ at strong electron-photon coupling, e.g., $\lambda=0.5\omega_0$,
	by comparing with the counterpart  based on the system Hamiltonian at Eq.~(\ref{hs1}).
	It is interesting to find that the photon number shows the aperiodic oscillation due to the effect of counter-rotating terms, which is significantly distinct from the one with the electron-photon interaction
	$(a^{\dagger}_{l}|g_{l}{\rangle}{\langle}e_l|
	+a_{l}|e_{l}{\rangle}{\langle}g_l|)$.
	Furthermore, the peak of photon number breaches the unity, which is unavailable in absence of counter-rotating terms.
	
	Next, we study the photon number dynamics at long time evolution in Fig.~\ref{fig3}(b).
	It is shown that by increasing the electron-photon coupling strength, the aperiodicity degree
	becomes increasingly apparent.
	Moreover, it is interesting to find that after long time evolution the photon number becomes
	nonvanishing.
	This directly demonstrates the breakdown of the complete measurement at Eq.~(\ref{cm1})
	with weak electron-photon interaction.
	We plot the inset of Fig.~\ref{fig3}(b) to quantify the robustness of the photon number.
	${\langle}a^\dag_{l}a_{l}{\rangle}(t{\rightarrow}\infty)$ becomes monotonic enhancement with the increase of $\lambda$.

	As is known that after long time evolution the system may reach the steady state,
	which generally may not be the eigenstate, e.g., the ground state at zero temperature.
	In particular for rather strong electron-photon coupling,
	by considering the degenerate perturbation theory,
	the eigenvectors of $H_{s,l}$ at Eq.~(\ref{hs2}) are estimated as
	\begin{eqnarray}
		|\psi_{n,\pm}{\rangle}&{\approx}&\frac{1}{\sqrt{2}}
		\Big(\exp\Big[{-\frac{\lambda}{\omega_0}(a^\dag_{l}-a_{l})}\Big]
		\frac{(a^\dag_{l})^n}{\sqrt{n!}}|0{\rangle}_a{\otimes}|-{\rangle}\nonumber\\
		&&{\pm}\exp\Big[{\frac{\lambda}{\omega_0}(a^\dag_{l}-a_{l})}\Big]
		\frac{(a^\dag_{l})^n}{\sqrt{n!}}|0{\rangle}_a{\otimes}|+{\rangle}\Big),
	\end{eqnarray}
	with $|\pm{\rangle}=(|e_{{l}}{\rangle}{\pm}|g_{{l}}{\rangle})/\sqrt{2}$
	and $a_{l}|0{\rangle}_a=0$.
	And the corresponding eigenvalues become
	$E_{n,\pm}\approx\omega_0n{\mp}|D_{nn}(2\lambda/\omega_0)|^2E_g/2  -\lambda^2/\omega_0+E_g/2$,
with the coherent state overlap coefficient
$D_{nn}=e^{-2(\lambda/\omega_0)^2}\sum^n_{r=0}(-1)^rn!(2\lambda/\omega_0)^{2n-2r}/[(n-r)!(n-r)!r!]$~\cite{qhchen2008pra}.
	Then, the transition rates between two eigenstate populations via the master equation could be reestablished.
	Specifically, by including the secular approximation
	the population dynamics is given by
	$dP_{n,\pm}/dt{\approx}[\kappa{(n+1)}P_{n+1,\pm}-\kappa{n}P_{n,\pm}]
	+\kappa(\lambda/\omega_0)^2(P_{n,\mp}-P_{n,\pm})$,
	with the density matrix elements
	$P_{n,\pm}={\langle}\psi_{n,\pm}|\rho_l|\psi_{n,\pm}{\rangle}$
	and the transition coefficients
	${\langle}\psi_{n-1,\pm}|a_{l}|\psi_{n,\pm}{\rangle}=\sqrt{n}$
	and
	${\langle}\psi_{n-1,\pm}|a_{l}|\psi_{n,\mp}{\rangle}=0$.
	The dissipative dynamics generally  results in the steady state
	\begin{eqnarray}~\label{ss1}
		\rho_{l}(\infty){\approx}\frac{1}{2}
		(|\psi_{0,+}{\rangle}{\langle}\psi_{0,+}|+|\psi_{0,-}{\rangle}{\langle}\psi_{0,-}|).
	\end{eqnarray}
	Thus, the corresponding steady-state photon number is given by
	${\langle}a^\dag_{l}a_{l}{\rangle}(t{\rightarrow}\infty){\approx}({\lambda}/{\omega_0})^2$.
	This partially explain the nonvanishing steady-state value of the average photon number at strong electron-photon coupling in the inset of Fig.~\ref{fig3}(b).
	Therefore, we conclude that strong electron-photon interaction generally breaks the microscopic particle conservation and  complete measurement.

	\subsection{Finite temperature reservoir}
	We study the effect of finite-temperature reservoir on the quantum measurement of probability.
	At weak electron-photon coupling $\lambda_v/\omega_{0,v}{\ll}1$,
	the master equation for dissipative dynamics of the electron-photon model can be described as
	\begin{eqnarray}
		\frac{d\rho_s}{dt}&=&-i\sum_{v={l},{r}}[H_{s,v},\rho_s]\\
		&&+\sum_{v={l},{r}}\kappa_v(1+\overline{n}_v)\mathcal{L}[a_v]\rho_s
		+\sum_{v={l},{r}}\kappa_v\overline{n}_v\mathcal{L}[a^\dag_v]\rho_s,\nonumber
	\end{eqnarray}
	where the dissipator denotes
	$\mathcal{L}[a_v]\rho_s=(
	2a_v\rho_sa^\dag_v-a^\dag_va_v\rho_s-\rho_sa^\dag_va_v)/2$.
	Here, we empirically describe the incoherent coupling between the photon resonator
	and the bosonic reservoir at the temperature $T_v$ with  the Bose-Einstein distribution function $\overline{n}_v=1/[\exp(\omega_{0,v}/k_BT_v)-1]$~\cite{jmfink2010prl,tbosen2023prl}.
	We select the initial state of the reduced $H_{s,v}$ as
	\begin{eqnarray}
		\rho_v(0)&=&(|C_v|^2|e_{v}{\rangle}{\langle}e_{v}|+|C_{\overline{v}}|^2|g_{v}{\rangle}{\langle}g_{v}|)\\
		&&{\otimes}
		\frac{e^{-\beta_v\omega_{0,v}{a^\dag}_va_v}}{\mathcal{Z}_v},\nonumber
	\end{eqnarray}
	with the partition function
	$\mathcal{Z}_v=1/[1-\exp(-\beta\omega_{0,v})]$
	and $\beta_v=1/k_BT_v$.

	We investigate the energy flow into the $v$-th reservoir.
	At low temperature regime, it is numerically found that the dissipative dynamics of the flow is typically contributed by three lowest eigenstates, i.e.,
	$|\psi_{1_v,\pm}{\rangle}$ and $|\psi_{0_v}{\rangle}$.
	Accordingly, relation between the energy flow and the photon number can be expressed as
\begin{eqnarray} ~\label{jvT}
J_v&\approx&\frac{(\omega_{0,v}-E_{g,v})}{2}
[\kappa_v(1+\overline{n}_v){\langle}a^\dag_v{a}_v{\rangle}-\kappa_v\overline{n}_vP_{0,v}]\nonumber\\
&&+\frac{\Lambda_v}{2}\frac{d(P_{-,v}-P_{+,v})}{dt},
\end{eqnarray}
	with $P_{\pm,v}={\langle}\psi_{1_v,\pm}|\rho_v|\psi_{1_v,\pm}\rangle$,
	$P_{0,v}={\langle}\psi_{0_v}|\rho_v|\psi_{0_v}\rangle$,
	and
	$\Lambda_v=\sqrt{(\omega_{0,v}+E_{g,v})^2+4\lambda^2_v}$.
Moreover, the total energy loss of the quantum system is obtained as
$\int^\infty_0J_vdt=[{\langle}H_{s,v}{\rangle}(0)-{\langle}H_{s,v}{\rangle}(\infty)]$,
which is approximated as
$\int^\infty_0J_vdt{\approx}E_{g,v}(1/[1+\exp(-\beta_vE_{g,v})]-|C_v|^2)$.
Hence, based on the microscopic energy conservation law,
	we obtain the effective probability measurement by integrating the current at Eq.~(\ref{jvT}) as
	\begin{eqnarray}~\label{pro-T}
    \kappa_v\int^\infty_0dt[{\langle}a^\dag_v{a}_v{\rangle}(t)-\frac{\overline{n}_vP_{0,v}(t)}{1+\overline{n}_v}]
		{\approx}\frac{|C_{{v}}|^2-1/\mathcal{Z}^\prime_v}
		{(1+\overline{n}_v)^2},
	\end{eqnarray}
with $\mathcal{Z}^\prime_v=(1+e^{-\beta_vE_{g,v}})$.
	If the temperature approaches zero, i.e., $\overline{n}_v=0$,
	the effective relation is naturally reduced to the complete probability measurement relation,
	i.e., $ \kappa_v\int^\infty_0dt{\langle}a^\dag_v{a}_v{\rangle}(t)=|C_v|^2$
	at Eq.~(\ref{cm1}).

	Alternatively, the dynamical equation of the ground state probability is described as
	\begin{eqnarray}~\label{dyn-T}
		\frac{d}{dt}P_{0,v}\approx\kappa(1+\overline{n}_v){\langle}a^\dag_v{a}_v{\rangle}
		-\kappa_v\overline{n}_vP_{0,v},
	\end{eqnarray}
	in which the change of the probability is contributed by
	the photon flow into the reservoir and the backward flow from the reservoir thermal noise.
	Straightforwardly, the effective probability measurement at Eq.~(\ref{pro-T}) can be recovered from Eq.~(\ref{dyn-T})
  	by considering
  $P_{0,v}(0){\approx}|C_{\overline{v}}|^2/\mathcal{Z}_v$
	and $P_{0,v}(\infty){\approx}1/\mathcal{Z}_v\mathcal{Z}^\prime_v$.

\section{Conclusion}
In summary, we investigate a practical measurement of the quantum probability and its underlying microscopic conservation laws through quantum electron-photon interaction and quantum photon dissipation, connecting the quantum probability with the probabilistic Born's rule. We model the measurement setup with two individual quantum dots, each with an electron occupation interacting with a single-mode photonic resonator, which is ultimately dissipated to a reservoir, e.g., a photon counter.

For the quantum measurement with weak electron-photon coupling, we first study the complete probability measurement at Eq.~\eqref{cm1}, which exactly connects the quantum probability of the electron wavefunction in the dot with the counted photon numbers. The measurement rule is protected by the microscopic energy conservation law with electron-photon energy exchange processes. Such rule can be alternatively recovered under the microscopic particle conservation law. This complete probability measurement relation is impervious to the quantum system parameters, such as $E_{e(g),v}$, $\omega_{0,v}$, and $\lambda_v$.
Next, we study the finite time probability measurement at finite time dynamics. Under the secular approximation, at resonance ($E_{e,v}-E_{g,v}=\omega_{0,v}$) the photon numbers at Eq.~(\ref{nt}) exhibit the periodic oscillating damping behavior, which are determined by the the electron-photon coupling and photon dissipation strengths. The emitted photon number shows intermittent increase feature at Eq.~(\ref{icm1}), with quasi-steps appearing around the times $t_v=n\pi/\lambda_v$. The characteristic measurement time is quantified as $t_{c,v}{\approx}2\pi/\kappa_v$, where the measurement outcome of emitted photons numbers and energy becomes effectively complete. Furthermore, it is found that the bias condition ($(E_{e,v}-E_{g,v}){\neq}\omega_{0,v}$) blocks the efficient measurement of the quantum probability.

	We further analyze the applicability of quantum measurement in the strongly-coupled electron-photon model and consider the counter-rotating terms in $H_{s,v}$ Eq.~(\ref{hs2}), which straightforwardly breaks the particle number conservation. In short-time dynamics, by comparing the photon number dynamics with the counterpart in absence of counter-rotating terms, we observe two distinct features:
	(1) the photon number exhibits the aperiodic oscillations, and
	(2) the peak of photon number breaches the unity.
	While in long time dynamics, it is found that
	(1) the aperiodicity degree of photon number oscillating behavior becomes apparent with the increase of electron-photon coupling, and
	(2) the photon number still becomes nonvanishing, due to the formation of the mixture of photonic coherent states, which  demonstrates the breakdown of the complete probability measurement relation.
	
Moreover, we investigate the influence of finite temperature reservoir on the quantum measurement and explore the effective measurement relation connecting the counted photon flux with quantum probability and thermal statistics.

We hope that the reported measurements  of quantum probability and the microscopic conservation laws may provide general avenues for practically detecting quantum probability.

\section{ACKNOWLEDGEMENTS}
This work was supported by the funding for the National Natural Science Foundation of China under Grants No. 12125504, No. 12074281, and No. 11704093, Jiangsu Key Disciplines of the Fourteenth Five-Year Plan (Grant No. 2021135), and the Opening Project of Shanghai Key Laboratory of Special Artificial Microstructure Materials and Technology.

\appendix*
\section{Derivation of complete quantum probability measurement relation}
In the eigenbasis of $H_{s,v}$, the energy current $J_v=-d{\langle}H_{s,v}{\rangle}/dt$ is expressed as
\begin{eqnarray}
J_v=-E_{+,v}\frac{dP_{v,+}(t)}{dt}-E_{-,v}\frac{dP_{v,-}(t)}{dt},
\end{eqnarray}
with the eigen-energies $E_{\pm,v}$ at Eq.~(\ref{Epm}).
Including the dynamics of density matrix operator in the eigenbasis of $H_{s,v}$
\begin{subequations}
\begin{align}
 \frac{d}{dt}P_{v,+}=&-\kappa\cos^2\frac{\theta}{2}P_{v,+}
 +\frac{\kappa}{2}\sin{{\theta}}\textrm{Re}[P_{v,+-}],\\
 \frac{d}{dt}P_{v,-}=&-\kappa\sin^2 \frac{\theta}{2}P_{v,-}+\frac{\kappa}{2}\sin{{\theta}}\textrm{Re}[P_{v,+-}],\\
\frac{d}{dt}P_{v,+-}=&-(i\Lambda_v+\frac{\kappa}{2})P_{v,+-}
+\frac{\kappa}{2}{\sin{\theta}}\textrm{Re}[P_{v,+-}],
\end{align}
\end{subequations}
with $\Lambda_v=E_{+,v}-E_{-,v}$ and $\textrm{Re}[P_{v,+-}]=(P_{v,+-}+P_{v,-+})/2$,
the current is specified as
\begin{eqnarray}
J_v&=&\frac{\kappa}{2}(2\cos^2\frac{\theta}{2}P_{v,+}
 -\sin{{\theta}}\textrm{Re}[P_{v,+-}])\nonumber\\
 &&+\frac{\kappa}{2}(2\sin^2\frac{\theta}{2}P_{v,-}
 -\sin{{\theta}}\textrm{Re}[P_{v,+-}]).
\end{eqnarray}
Considering the expression of photon number ${\langle}a^\dag_la_l{\rangle}$ at Eq.~(\ref{nt1}), the relation between the current and the photon number can be established as
\begin{eqnarray}~\label{app-jv}
J_v&=&\frac{\kappa}{2}(\omega_{0,v}-E_{g,v}){\langle}a^\dag_la_l{\rangle}\\
&&-\frac{\Lambda_v}{2}
\Big(\frac{d}{dt}P_{v,+}-\frac{d}{dt}P_{v,-}\Big).\nonumber
\end{eqnarray}
By integrating Eq.~(\ref{app-jv}) at both sides, we find
$\int^\infty_0J_vdt=-E_{g,v}|C_v|^2$, $P_{v,-}(0)-P_{v,+}(0)=|C_v|^2\cos\theta$,
	and $P_{v,-}(\infty)-P_{v,+}(\infty)=0$.
Finally, we obtain the complete quantum measurement relation at Eq.~(\ref{cm1}).


\begin{thebibliography}{99}
		
		\bibitem{sw1989ap} S. Weinberg,
		%Testing quantum mechanics,
		Ann. Phys. (NY) \textbf{194}, 336 (1989).
		\bibitem{sw1989prl}
		S. Weinberg,
		%Precision Tests of Quantum Mechanics,
		Phys. Rev. Lett. \textbf{62}, 485 (1989).
		
		
		\bibitem{mborn1926zp}
		M. Born,
		%Quantenmechanik der Stovorgange,
		Z. Phys. \textbf{38}, 803 (1926).
		\bibitem{mborn1954nl}
		M. Born, The statistical interpretation of quantum mechanics, Nobel Lecture, (1954).
		\bibitem{usinha2010science}
		U. Sinha, C. Couteau, T. Jennewein, R. Laflamme, and G. Weihs, %Ruling Out Multi-Order Interference in Quantum Mechanics,
		Science \textbf{329}, 418 (2010).
		\bibitem{abassi2013rmp}
		A. Bassi, K. Lochan, S. Satin, T. P. Singh, and H. Ulbricht,
		%Models of wave-function collapse, underlying theories, and experimental tests,
		Rev. Mod. Phys. \textbf{85}, 471 (2013).
		\bibitem{pshadbolt2014np}
		P. Shadbolt, J. C. F. Mathews, A. Laing, and Jeremy L. O'Brien,
		%Testing foundations of quantum mechanics with photons,
		Nature Physics \textbf{10}, 278 (2014).
		\bibitem{mop2021prl}
		M. O. Pleinert, A. Rueda, E. Lutz, and J. von Zanthier,
		%Testing Higher-Order Quantum Interference with Many-Particle States
		Phys. Rev. Lett. \textbf{126}, 190401  (2021).
		
		
		\bibitem{jaw1984book}
		J. A. Wheeler and W. H. Zurek,  \emph{Quantum Theory and Measurement}
		(Princeton University Press, 2014).
		\bibitem{hmwiseman2011book}
		H. M. Wiseman and G. J. Milburn,
		\emph{Quantum Measurement and Control}
		(Cambridge University Press, Cambridge, 2011).

        \bibitem{aac2010rmp}
        A. A. Clerk, M. H. Devoret, S. M. Girvin, Florian Marquardt, and R. J. Schoelkopf, Rev. Mod. Phys. \textbf{82}, 1155 (2010).p		

		\bibitem{whz2003prl}
		W. H. Zurek,
		%Environment-Assisted Invariance, Entanglement, and Probabilities in Quantum Physics,
		Phys. Rev. Lett. \textbf{90}, 120404  (2003).
		\bibitem{whz2005pra}
		W. H. Zurek,
		%Probabilities from entanglement, Born's rule $p_k=|\psi_k|^2$ from envariance,
		Phys. Rev. A \textbf{71}, 052105 (2005).
		
		\bibitem{ms2005fop}
		M. Schlosshauer and A. Fine,
		%On Zurek's Derivation of the Born Rule,
		Foundations of Physics  \textbf{35}, 197 (2005).
		\bibitem{um2004ijqi}
		U. Mohrhoff,
		%Probabilities from envariance?,
		Int. J. Quantum Inform. \textbf{2}, 221 (2004).
		
		
		\bibitem{aea2013rmp}
		A. E. Allahverdyana, R. Balianb, and T. M. Nieuwenhuizenc,
		%Understanding quantum measurement from the solution of dynamical models,
		Phys. Rep. \textbf{525}, 1 (2013).
		
		\bibitem{bs2001book}
		J. S. Bell, \emph{On wave packet reduction in the Coleman-Hepp model. In John S Bell On The Foundations Of Quantum Mechanics}
		(World Scientific, Singapore, 2001).
		\bibitem{gb1990jsp}
		B. Gaveau and L. Schulman,
		%Model apparatus for quantum measurements,
		J. Stat. Phys.  \textbf{58}, 1209 (1990).
		
		\bibitem{aea2003epl}
		A. E. Allahverdyan, R. Balian, and Th. M. Nieuwenhuizen,
		%Curie-Weiss model of the quantum measurement process,
		Europhys. Lett. \textbf{61},  452 (2003).
		
		\bibitem{mrd2011prl}
		M. R. Delbecq, V. Schmitt, F. D. Parmentier, N. Roch, J. J. Viennot, G. Feve, B. Huard, C. Mora, A. Cottet, and T. Kontos,
		%Coupling a Quantum Dot, Fermionic Leads, and a Microwave Cavity on a Chip,
		Phys. Rev. Lett. \textbf{107}, 256804 (2011).
		
		\bibitem{yyliu2014prl}
		Y. Y. Liu, K. D. Petersson, J. Stehlik, J. M. Taylor, and J. R. Petta,
		%Photon Emission from a Cavity-Coupled Double Quantum Dot,
		Phys. Rev. Lett. \textbf{113}, 036801 (2014).
		
		\bibitem{xmi2017science}
		X. Mi, J. V. Cady, D. M. Zajac, P. W. Deelman, and J. R. Petta,
		%Strong coupling of a single electron in silicon to a microwave photon,
		Science \textbf{355}, 156 (2017).
		
		\bibitem{trh2018prl}
		T R. Hartke, Y.-Y. Liu, M. J. Gullans, and J. R. Petta,
		%Microwave Detection of Electron-Phonon Interactions in a Cavity-Coupled Double Quantum Dot,
		Phys. Rev. Lett. \textbf{120}, 097701  (2018).
		
        \bibitem{jhjiang2014prx}
        J. H. Jiang and S. John, Phys. Rev. X \textbf{4}, 031025 (2014).

        \bibitem{rqwang2022adx}
        R. Q. Wang, C. Wang, J. C. Lu, and J. H. Jiang, Advances in Physics:X \textbf{7}, 2082317 (2022).		

		\bibitem{hpb2002book}
		H. J. Carmichael,
		\emph{Statistical Methods in Quantum Optics 1:
			Master Equations and Fokker-Planck Equations}  (Springer-Verlag, New York, 1999).
		\bibitem {uweiss2012book}
		U. Weiss, \emph{Quantum Dissipative Systems} (World Scientific, Singapore, 2012).
		
		
		\bibitem{syzhu1988pla}
		S. Y. Zhu, Z. D. Liu, and X. S. Li,
		%Squeezing in a three-level Jaynes-Cummings model,
		Phys. Lett. A \textbf{128}, 89 (1988).
		\bibitem{vva2012prl}
		V. V. Albert,
		%Quantum Rabi Model for  $N$-State Atoms,
		Phys. Rev. Lett. \textbf{108}, 180401  (2012).
		\bibitem{pstegman2022acs}
		P.  Stegman, S. N. Gupta, G. Haran, and J. S. Cao,
		%Higher-Order Photon Statistics as a New Tool to Reveal Hidden Excited States in a Plasmonic Cavity,
		ACS Photonics \textbf{9}, 2119 (2022).
		
		
		\bibitem{rl2020prx}
		R. Lescanne, S. Deleglise, E. Albertinale, U. Reglade, T. Capelle, E. Ivanov, T. Jacqmin, Z. Leghtas, and E. Flurin,
		%Irreversible Qubit-Photon Coupling for the Detection of Itinerant Microwave Photons,
		Phys. Rev. X \textbf{10}, 021038 (2020).
		\bibitem{ea2021nature}
		E. Albertinale, L. Balembois, E. Billaud, V. Ranjan, D. Flanigan, T. Schenkel, D. Esteve, D. Vion, P. Bertet, and E. Flurin,
		%Detecting spins by their fluorescence with a microwave photon counter,
		Nature \textbf{600}, 434 (2021).
		
\bibitem{man2002book}
M. A. Nielsen and I. Chuang, \emph{Quantum Computation and
Quantum Information} (Cambridge University Press, 2010).
\bibitem{jcorler2020prl}
J. Cotler and F. Wilczek, Phys. Rev. Lett. \textbf{124}, 100401 (2020).

        \bibitem{note1}
        Actually, there exists slight increse around each quasi-step (not exact step) of countered photon numbers, which is contributed by negligible photon flow around $nT$.

		\bibitem{mk2014prb}
		M. Kulkarni, O. Cotlet, and H. E. Tureci,
		%Cavity-coupled double-quantum dot at finite bias: Analogy with lasers and beyond,
		Phys. Rev. B \textbf{90}, 125402 (2014).
		\bibitem{bka2016prb}
		B. K. Agarwalla, M. Kulkarni, S. Mukamel, and D. Segal,
		%Giant photon gain in large-scale quantum dot-circuit QED systems,
		Phys. Rev. B \textbf{94}, 121305(R) (2016).
		\bibitem{jclu2019prb}
		J. C. Lu, R. Q. Wang, J. Ren, M. Kulkarni, and J. H. Jiang,
		%Quantum-dot circuit-QED thermoelectric diodes and transistors
		Phys. Rev. B \textbf{99}, 035129 (2019).
		
		\bibitem{dbraak2011prl}
		D. Braak,
		%Integrability of the Rabi Model,
		Phys. Rev. Lett. \textbf{107}, 100401  (2011).
		\bibitem{afk2019nrp}
		A. F. Kockum, A. Miranowicz, S. De Liberato, S. Savasta, and F. Nori,
		%Ultrastrong coupling between light and matter,
		Nature Reviews Physics \textbf{1}, 19 (2019).
		\bibitem{pfd2019rmp}
		P. Forn-Diaz, L. Lamata, E. Rico, J. Kono, and E. Solano,
		%Ultrastrong coupling regimes of light-matter interaction,
		Rev. Mod. Phys. \textbf{91}, 025005 (2019).
		
		\bibitem{qhchen2012pra}
		Q. H. Chen, C. Wang, S. He, T. Liu, and K. L. Wang,
		%Exact solvability of the quantum Rabi model using Bogoliubov operators,
		Phys. Rev. A \textbf{86}, 023822 (2012).
		
        \bibitem{qhchen2008pra}
        Q. H. Chen, Y. Y. Zhang, T. Liu, and K. L. Wang, Phys. Rev. A \textbf{78}, 051801(R) (2008).

		\bibitem{jmfink2010prl}
		J. M. Fink, L. Steffen, P. Studer, Lev S. Bishop, M. Baur, R. Bianchetti, D. Bozyigit, C. Lang, S. Filipp, P. J. Leek, and A. Wallraff,
		%Quantum-To-Classical Transition in Cavity Quantum Electrodynamics,
		Phys. Rev. Lett. \textbf{105}, 163601 (2010).
		\bibitem{tbosen2023prl}
		T. Bonsen, P. Harvey-Collard, M. Russ, J. Dijkema, A. Sammak, G. Scappucci,
		and L. M. K. Vandersypen,
		%Probing the Jaynes-Cummings Ladder with Spin Circuit Quantum Electrodynamics,
		Phys. Rev. Lett. \textbf{130}, 137001 (2023).
			
	\end{thebibliography}
\end{document}